\documentclass[11pt]{article}

\usepackage[utf8]{inputenc}
\usepackage[T1]{fontenc}
\usepackage{lmodern}
\usepackage{microtype}
\usepackage[margin=1in]{geometry}

\usepackage{amsmath}
\usepackage{amssymb}
\usepackage{amsthm}

\usepackage{graphicx}
\usepackage{booktabs}

\usepackage[round,authoryear,sort]{natbib}

\usepackage[colorlinks=true,allcolors=blue,breaklinks=true]{hyperref}

\theoremstyle{plain}

\theoremstyle{definition}

\newcommand{\Eb}{\mathbb{E}}
\newcommand{\Nrm}[2]{\mathcal{N}\!\left(#1,#2\right)}
\newcommand{\indic}[1]{\mathbf{1}\!\left[#1\right]}
\newcommand{\RPS}{\mathrm{RPS}}
\newcommand{\BS}{\mathrm{BS}}

\title{Pre-game paired-comparison modeling of professional\\
       \emph{League of Legends} map outcomes}

\author{%
  Min-Ren Guan\thanks{Teahouse Finance, Taipei 105803, Taiwan.
    E-mail: \texttt{john@teahouse.finance}.}
  \and
  Shen-Ning Tung\thanks{Corresponding author. Department of Mathematics,
    National Tsing Hua University, No.~101, Sec.~2, Guangfu Rd., Hsinchu
    300044, Taiwan. E-mail: \texttt{tung@math.nthu.edu.tw}.
    ORCID: 0000-0002-4646-3575.}%
}

\date{December 1, 2026}

\begin{document}

\maketitle

\begin{abstract}
We build and evaluate a pre-game win-probability forecaster for individual
maps (``games'') in professional \emph{League of Legends} (LoL). The proposed model is a one-stage
logistic regression with three components: each team's
exponentially-weighted moving average of past same-side results, a
ridge-shrunk stable strength that is the maximum-a-posteriori estimate of a
logistic mixed model, and a first-pick draft covariate. Its probabilities come directly from the
fitted model, with no separate calibration step (calibration slope $0.978$ in
the walk-forward and $0.81$ on the global holdout). It improves on a purely
dynamic Bradley--Terry specification by adding stable team strengths. A second, independently built two-stage composite
mixed model under restricted maximum likelihood (REML) and best linear
unbiased prediction (BLUP) shrinkage, with Platt calibration, serves as
the most complete two-stage rival evaluated here.
On $5{,}135$ games across six regional leagues and three international
events (2024--2026), under paired per-game Diebold--Mariano inference, the
two architectures are statistically indistinguishable on every reported
cross-region protocol and window (global holdout $0.2261$ vs.\ $0.2274$;
walk-forward $0.2208$ vs.\ $0.2210$), so the simpler model is preferred on
parsimony, not accuracy. Both score below the static fits ($0.2326$/$0.2286$)
and the classical dynamic benchmark ($0.2382$). Against Polymarket on $907$ of
$912$ matched maps, the market is ahead even on per-map contracts; among league
and event slices, only cross-region Worlds shows a gap that survives Bonferroni
correction.
\end{abstract}


\bigskip


\section{Introduction} \label{sec:intro}

Competitive esports has, over the past decade, matured into a data-rich analogue of the professional sports leagues that have long been the primary subject matter of quantitative sports analysis. Matches are scheduled, officiated, and broadcast under standardized formats, and, increasingly, liquid prediction markets price the same contests, giving a statistical forecaster an external benchmark to be judged against, beyond classical goodness-of-fit metrics. \emph{League of Legends} (\emph{LoL}) is among the most heavily professionalized and extensively documented of these titles, contested in regional leagues across China (LPL), Korea (LCK), Europe (LEC), North America (LCS), the Asia-Pacific (LCP), and Brazil (CBLOL), together with recurring international championships (Worlds, the Mid-Season Invitational, and the First Stand event). A professional match consists of one or more \textbf{maps}---individual games, each ending in a clean win or loss---most commonly contested as a \textbf{best-of-three} (\textbf{Bo3}), the series awarded to the first team to win two maps. Before each map begins, the two teams alternate selecting and banning playable characters in a structured \textbf{draft}; the team that picks first in that phase holds \textbf{first-pick} rights, a draft-order covariate rather than a fixed property of either team. Before 2026 the first pick went with the blue side; from 2026, under Riot's First Selection rules, the team holding the right of first selection chooses the side or the draft order, and its opponent chooses the remaining option \citep{riot2026firstselection}.

Quantitative work specific to esports win prediction remains comparatively young next to the century-old paired-comparison literature it draws on, and has so far concentrated on two distinct problems: forecasting from rich in-game or player-level features, and forecasting from the pre-game information available before a single character is played. \citet{maymin2021smart} develops kill- and death-level analytics for LoL that connect individual in-game events to win probability, establishing that the game's telemetry carries exploitable predictive signal well beyond the final score. \citet{hodge2021win} build a live, in-game win-prediction system evaluated on professional Dota~2 matches, demonstrating that continuously updated in-play features substantially outperform static pre-match priors as a contest unfolds---a live-model problem the present paper deliberately does not address, restricting itself throughout to information available before the map begins. Closest in spirit to the present exercise, \citet{jadowski2022statistical} fit statistical classifiers to metrics gathered from professional LoL's 2020 international competitions, their best specification correctly calling $67\%$ of the $306$ games evaluated; their study, however, neither adopts a dynamic paired-comparison architecture nor benchmarks against a prediction market, leaving both as open contributions this paper takes up directly. Read together, this literature motivates evaluating pre-game forecasts as a question worth answering in its own right, distinct from the in-game prediction problem, while leaving unresolved how much modeling complexity that pre-game forecasting task actually requires and how the resulting forecasts compare with the market's own pricing---the two questions this paper is built to answer.

Three research questions organize the analysis. First, does a deliberately minimal one-stage logistic model---a small fixed-coefficient block plus ridge-shrunk team strengths, fit end-to-end on the forecasting loss it is judged by---concede anything in held-out accuracy to the most complete two-stage rival evaluated here, the composite mixed model of \S\ref{sec:twostage}, evaluated under the same information and scoring protocols? Second, do either of these two architectures improve on the classical paired-comparison benchmarks against which any new dynamic model should be measured---Cattelan, Varin, and Firth's purely dynamic Bradley--Terry model and a static, Stefani-style least-squares fit? Third, does the resulting pre-game forecaster price individual maps as accurately as Polymarket, a liquid prediction market pricing contracts on the same games? The two architectures prove statistically indistinguishable on every reported cross-region protocol and window, so the case for the simpler model rests on its needing no composite response and no post-hoc calibration step; the market is ahead on the matched per-map contracts.

The paper's models sit in a direct methodological lineage. \citet{stefani1977football} and \citet{harville1977use} established static paired-comparison rating from margins of victory, assigning each competitor a fixed strength recovered by least squares; such models cannot represent within-season form, a limitation acute in a domain where rosters change between splits and the underlying game itself is periodically rebalanced by a \textbf{patch}, a scheduled software update to character and item statistics. Cattelan, Varin, and Firth (\citeyear{cattelan2013dynamic}) resolved the static model's chief limitation with a purely dynamic Bradley--Terry model in which team ability follows an autoregressive, exponentially-weighted process driven entirely by past results and carries no stable per-team component of its own, continuing a state-space paired-comparison tradition developed by \citet{barry1993choice}, \citet{glickman1999parameter}, and \citet{fahrmeir1994dynamic} that this paper does not otherwise draw on directly. A separate line, following the same dynamic idea but replacing the binary outcome with a continuous score or margin composite under a linear mixed model, was developed by \citet{guan2020analyzing} in an unpublished master's thesis, worked there on professional basketball; the present paper is, to our knowledge, the first application of that continuous-response, two-stage paired-comparison design to esports---a domain the line had not previously reached---and carries it forward as the second candidate model of \S\ref{sec:twostage}. Against this lineage, the paper's own contributions are four: a one-stage model that completes Cattelan, Varin, and Firth's purely dynamic Bradley--Terry with the stable team strength it structurally lacks, in a single estimation pass rather than two, established here as exactly the maximum-a-posteriori solution of a logistic mixed model under a ridge penalty---an equivalence that is this paper's own construction rather than one carried from either source; a first-pick covariate for LoL's draft order, coded so that it is estimated only where the draft rule varies (from 2026); a benchmark against Polymarket (\S\ref{sec:market}); and, as a secondary numerical point, that the one-stage model's ridge penalty makes its team strengths identifiable by construction, whereas the classical random-effect fit is identified only through a restricted-likelihood optimization whose result must be checked, since one prespecified optimizer reports convergence at points that are not maxima, on the full corpus as on single leagues (\S\ref{sec:workflow}).

The remainder of the paper proceeds in the order the model itself motivates. Section~\ref{sec:model} specifies the paired-comparison lineage, the proposed one-stage model, and the second candidate two-stage model, establishing the target quantity---the pre-game win probability $\widehat p_i$---before any data are described. Section~\ref{sec:data} describes the data sources feeding the corpus and the market benchmark, the resulting scope and the source properties that shape the ingestion rules, and the two complementary holdout protocols used throughout. Section~\ref{sec:rps} defines the ranked probability score and its exact reduction to the Brier score for a two-outcome per-map market, together with the paired standard error appropriate to every reported comparison. Section~\ref{sec:results} reports the results, moving from the classical benchmarks through the one-stage/two-stage architecture comparison and its ablations to the market benchmark. Section~\ref{sec:conclusion} concludes.


\section{Model} \label{sec:model}

The quantity this paper delivers and scores throughout is the pre-game win probability $\widehat p_i$ for map $i$'s per-map market: every specification decision below is judged by held-out forecast loss (\S\ref{sec:rps}) and, ultimately, against the prediction market itself (\S\ref{sec:market}). The \textbf{proposed model} (\S\ref{sec:proposed}) is a one-stage outcome logistic model: a paired-comparison forecaster fit to win/loss outcomes alone. The \textbf{second candidate model} (\S\ref{sec:twostage}) implements the two-stage rating tradition in full. Both are validated against each other and against two classical external benchmarks, dynamic and static (\S\ref{sec:cattelan}). Cross-region pooling is used throughout, and no patch fixed effect is fit. Throughout, map $i$ is played between blue-side team $b(i)$ and red-side team $r(i)$ on patch $p(i)$ at calendar time $t_i$; $O_i \in \{0,1\}$ is the blue-side win indicator, and $Y_i$ denotes a blue-perspective margin response wherever a layer of the road map below uses one.

\subsection{Road map: the paired-comparison lineage} \label{sec:roadmap}

The modeling landscape is best read as a short lineage, each step adding one piece of structure and answering one question that the previous step left open. The two layers this paper builds on most directly are stated in full below rather than merely cited; the two-stage line that historically follows is summarized at the end of the road map, and its most complete instantiation evaluated here is carried forward as the second candidate model (\S\ref{sec:twostage}).

\textbf{Layer 1: static paired comparison.} The base layer treats the score difference of a contest as a difference of latent, time-invariant team strengths. \citet{stefani1977football} assigns \emph{one} least-squares rating per team, fit to win margins. The blue/red asymmetry of the present domain makes a side-specific pair the natural adaptation of that design: writing $a^{\mathrm B}_{b(i)}$ for the blue team's strength and $a^{\mathrm R}_{r(i)}$ for the red team's, the least-squares model fitted here is
\begin{equation}
\label{eq:stefani}
Y_i \;=\; a^{\mathrm B}_{b(i)} - a^{\mathrm R}_{r(i)} + \varepsilon_i,
\qquad \varepsilon_i\sim\Nrm{0}{\sigma^2}.
\end{equation}
This side-specific design carries exactly one redundancy---shifting both blocks of strengths by a common constant leaves every prediction unchanged---so that $2(T-1)$ relative strengths together with a single blue$-$red \emph{level} remain identifiable, that level playing precisely the side-advantage role Stefani's original omits. \citet{harville1977use} gives the same idea its mixed-model form, with teams entering as random effects fitted to score margins and the strengths permitted to drift \emph{between} seasons through a first-order autoregression---an early formal treatment of time-varying strength that predates the state-space paired-comparison line discussed below. What a static layer of this kind buys is a coherent, transitive strength ranking, and, on this corpus, a lower held-out score than the purely dynamic Layer~2 that follows it (\S\ref{sec:cattelan}); what it cannot answer is anything about within-season \emph{form}, since a team's strength is a single number for the whole sample---Harville's autoregression moves it between seasons, not within them---so a mid-season hot or cold streak is invisible to it.

\textbf{Layer 2: canonical dynamic Bradley--Terry.} The second layer makes ability move \emph{within} the season but discards stable strength entirely. The Bradley--Terry model of \citet{bradley1952rank} predicts a pairwise win from the difference of two latent team strengths, the same difference-of-strengths logic that \citet{elo1978rating} put to practical use in chess rating by updating each player's strength after every result, and the state-space treatments of time-varying ability cited in \S\ref{sec:intro} filter those strengths from observed results. Cattelan, Varin, and Firth (\citeyear{cattelan2013dynamic}) replace this state-space machinery with a designed covariate: they model the binary outcome $O_i \in \{0,1\}$ with a logistic link and let each team's ability be \emph{purely dynamic}---a running form score updated after every game by an exponentially-weighted moving average (\textbf{EWMA}) of its own past results, with no underlying time-invariant component at all. Writing $S^{\mathrm B}_{b(i)}(t_i)$ for the blue team's purely dynamic same-side win form entering map $i$, and $t^{-}$ for the time of team $b(i)$'s previous blue-side game, the form evolves by the EWMA recursion, with blue-side smoothing parameter $\lambda_1 \in [0,1]$,
\begin{equation}
\label{eq:cattelan-ewma}
S^{\mathrm B}_{b(i)}(t_i) \;=\; (1-\lambda_1)\, S^{\mathrm B}_{b(i)}(t^{-})
  \;+\; \lambda_1\, O(t^{-}),
\end{equation}
where $O(t^{-}) \in \{0,1\}$ is the blue-perspective outcome of that previous blue-side game---team $b(i)$ played blue there, so it is exactly its own win indicator---and the sequence is initialized at the league-wide mean blue-side win rate of the previous season, common to all teams. The red-side form updates symmetrically from team $r(i)$'s previous red-side game at $t'^{-}$, whose win indicator for the red team is $1 - O(t'^{-})$, with its own smoothing parameter $\lambda_2$:
\begin{equation}
\label{eq:cattelan-ewma-red}
S^{\mathrm R}_{r(i)}(t_i) \;=\; (1-\lambda_2)\, S^{\mathrm R}_{r(i)}(t'^{-})
  \;+\; \lambda_2\,\bigl(1-O(t'^{-})\bigr).
\end{equation}
There is no stable component in this layer; the form \emph{is} the ability. The win probability is the logistic transform of the contemporaneous form gap,
\begin{equation}
\label{eq:cattelan-lin}
\Pr(O_i = 1) \;=\;
  \frac{\exp\!\bigl\{\gamma_1 S^{\mathrm B}_{b(i)}(t_i) - \gamma_2 S^{\mathrm R}_{r(i)}(t_i)\bigr\}}
       {1+\exp\!\bigl\{\gamma_1 S^{\mathrm B}_{b(i)}(t_i) - \gamma_2 S^{\mathrm R}_{r(i)}(t_i)\bigr\}} .
\end{equation}
In words, the probability that blue wins map $i$ rises with blue's own recent same-side form and falls with red's, on a logistic scale, and with no additive contribution from either team's identity beyond that recent record. The side-specific scales $\gamma_1, \gamma_2$ are estimated by conditional likelihood rather than fixed a priori, and teams are initialized at a common prior-season average. Cattelan, Varin, and Firth (\citeyear{cattelan2013dynamic}) treat the decays as \emph{estimated} parameters---per-side nuisance smoothing parameters $(\lambda_1, \lambda_2)$, fit by two-step maximum profile likelihood, with $\widehat\lambda_1 = 0.043$ and $\widehat\lambda_2 = 0.025$ in their National Basketball Association application, four free parameters in all---whereas the re-implementation of \S\ref{sec:cattelan} selects a single decay on a grid. What this layer buys is responsiveness to recent form, the headline strength of the dynamic Bradley--Terry family; what it gives up is a persistent identity for each team, since two clubs with identical recent records are statistically indistinguishable to it even if one has been the stronger roster all year. Whether that omission costs anything in forecast accuracy is an empirical question, answered in \S\ref{sec:cattelan}.

One point is worth stating explicitly: Layers~1 and~2 are \emph{not} boundary cases of one another. Freezing the decays at $\lambda_1 = \lambda_2 = 0$ in \eqref{eq:cattelan-ewma}--\eqref{eq:cattelan-ewma-red} does not recover per-team static strengths---it yields an intercept-only model, since Cattelan, Varin, and Firth's EWMA has no per-team stable component at any decay. Both layers are instead boundary cases of the second candidate model developed in \S\ref{sec:twostage}. The two-stage tradition that historically follows Layer~2 (\S\ref{sec:intro}) survives here as the second candidate model of \S\ref{sec:twostage}.

\subsection{The one-stage family and the proposed model} \label{sec:proposed}

The model this paper proposes is the minimal member of a one-stage family, fit end-to-end on the deliverable, that completes the road map's Layer~2 with the stable strength it lacks. A member of the family is indexed by the dimension $d$ of the game summary its latent update consumes. Summarize each finished game, from the perspective of the side that played it, by a feature vector $x \in \mathbb{R}^d$---directional own-minus-opponent quantities, sign-flipped for red, with game duration (the one undirected feature in the pool) entering unsigned---and let each team carry side-specific form states $E^{\mathrm B}, E^{\mathrm R} \in \mathbb{R}^d$, the coordinate-wise vector analogue of the Layer-2 recursion \eqref{eq:cattelan-ewma}: with $t^{-}$ the time of team $b(i)$'s previous blue-side game and $x(t^{-})$ that game's summary from its perspective,
\begin{equation}
\label{eq:onestage-ewma}
E^{\mathrm B}_{b(i)}(t_i) \;=\; (1-\lambda)\, E^{\mathrm B}_{b(i)}(t^{-})
  \;+\; \lambda\, x(t^{-}),
\qquad E^{\mathrm B}_{b(i)}(0) = \mathbf{0},
\end{equation}
zero-initialized at a team's competitive debut, with decay $\lambda$ selected on a grid; the red-side analogue updates $E^{\mathrm R}_{r(i)}$ from team $r(i)$'s previous red-side game in the same way. In words, \eqref{eq:onestage-ewma} tracks each team's recent form separately for the side it played on, weighting its most recent games most heavily and discounting older ones geometrically, exactly as \eqref{eq:cattelan-ewma} does for the scalar win indicator but now for a general $d$-dimensional game summary. The win probability is a single logistic expression,
\begin{equation}
\label{eq:onestage}
p_i \;=\; \operatorname{expit}\bigl(\beta_0
  + \gamma_{\mathrm{FP}}\,(\indic{\text{blue is first-pick}_i}-1)
  + \beta_B^{\top} E^{\mathrm B}_{b(i)}(t_i)
  - \beta_R^{\top} E^{\mathrm R}_{r(i)}(t_i)
  + \theta_{b(i)} - \theta_{r(i)}\bigr),
\end{equation}
where $\operatorname{expit}(x) = e^{x}/(1+e^{x})$, with side slope vectors $\beta_B, \beta_R \in \mathbb{R}^d$ and one stable strength $\theta_t$ per team. Three glosses on \eqref{eq:onestage} matter for what follows. \textbf{First pick} is professional LoL's draft-order covariate: one team picks first in the pre-game pick/ban phase, in which teams choose the \textbf{champions}---the playable characters---each of their players will use. Before 2026 the first pick went with the blue side. From 2026, under Riot's First Selection rules, the team holding the right of first selection chooses the side or the draft order, and its opponent chooses the remaining option \citep{riot2026firstselection}; in the LEC, for games after the first in a series, the right goes to the loser of the previous game \citep[\S6.8.5]{lec2026rulebook}. The indicator is coded relative to blue first pick, so the term is zero before 2026. Because the first pick is a team's choice, $\gamma_{\mathrm{FP}}$ measures a conditional association between picking first and winning, given the side, the teams and their form, not a causal effect of draft order. The coefficient is estimated only when red-first games appear among the training rows, and is otherwise fixed at zero; where it is estimated it carries the Gaussian prior given below. In particular, the global holdout's training window ends in 2025 and estimates no first-pick coefficient; among the within-league windows of Table~\ref{tab:perleague} only LCP's contains red-first games, and in the walk-forward the term first enters on 14 January 2026. And each side's form state consumes only that side's own games---a team's red-side results never update its blue-side form; the split is inherited from Layer~2, where it carries a real asymmetry that the model's intercepts absorb, and is kept throughout for comparability across every model in the road map. Pooling both sides into a single state was not tried. Finally, the features entering \eqref{eq:onestage-ewma} are normalized on training rows alone (\S\ref{sec:workflow}), with the directional margins scaled by their training-window standard deviations but deliberately \emph{not} centered---centering would destroy the within-game antisymmetry $x_{\text{team}} = -x_{\text{opponent}}$ that the sign-flipped construction guarantees---while game duration, having no direction to preserve, is fully standardized to zero mean and unit variance, and the win indicator is already coded on the symmetric $\pm 1$ scale. Because the slope vectors $\beta_B, \beta_R$ are left unpenalized, the fitted model is invariant to any per-feature rescaling of this kind, so the feature-count ablation of \S\ref{sec:onestage} cannot depend on the scaling choice.

The \emph{proposed} model is the minimal member of the family, $d=1$: the game summary is the signed result $x = 2O(t^{-}) - 1 \in \{\pm 1\}$ alone, so $E^{\mathrm B}_{b(i)}(t_i)$ is simply team $b(i)$'s EWMA of its own past blue-side results, and \eqref{eq:onestage} specializes to the scalar-slope expression
\begin{equation}
\label{eq:proposed}
p_i \;=\; \operatorname{expit}\bigl(\beta_0
  + \gamma_{\mathrm{FP}}\,(\indic{\text{blue is first-pick}_i}-1)
  + \beta_B\, E^{\mathrm B}_{b(i)}(t_i) - \beta_R\, E^{\mathrm R}_{r(i)}(t_i)
  + \theta_{b(i)} - \theta_{r(i)}\bigr)
\end{equation}
---four coefficients plus the shrunk team block. Put plainly, \eqref{eq:proposed} is a logistic regression that prevents over-fitting by shrinking each team's rating toward zero, the average over all teams in the fit, using a team's recent momentum---weighted toward its most recent games on the same side---and a first-pick indicator. The larger members of the family ($d=4$: the signed result plus the two-stage model's three margins; $d=10$: every end-of-game feature considered) are fit under an identical protocol in \S\ref{sec:onestage}; the $d$-ablation there finds no statistically clear accuracy difference among them on the holdout, so $d=1$ is chosen on parsimony.

Every member of the family is estimated in a single pass by minimizing the penalized log-loss
\begin{equation}
\label{eq:proposed-loss}
-\sum_i \bigl[\, O_i \log p_i + (1-O_i)\log(1-p_i) \,\bigr]
  \;+\; \frac{\tau}{2}\,\lVert\theta\rVert^2
  \;+\; \frac{\omega}{2}\,\gamma_{\mathrm{FP}}^2 ,
\end{equation}
where $\theta$ stacks the stable per-team strengths of \eqref{eq:onestage} into one vector and $\lVert\theta\rVert^2 = \sum_t \theta_t^2$ is its squared Euclidean norm, so the first penalty is simply the sum of squared team strengths, one term per team. The second penalty, with $\omega = 4$ fixed a priori, is the MAP form of the prior $\gamma_{\mathrm{FP}} \sim \Nrm{0}{0.5^2}$ on the log-odds scale: at $p = 0.5$, where one game carries the most information, it is worth about sixteen red-first games with no first-pick effect, which keeps the coefficient finite when only a few red-first games are in training while leaving it free once many are. The intercept and the slopes $(\beta_0, \beta_B, \beta_R)$---$1+2d$ numbers, three for the proposed member---are left unpenalized. The ridge term in \eqref{eq:proposed-loss} is not a tuning afterthought but the source of the model's principal methodological claim: the minimizer of \eqref{eq:proposed-loss} is exactly the \textbf{maximum-a-posteriori} (MAP) estimate of a logistic generalized linear mixed model with $\theta_t \sim \Nrm{0}{\sigma_\theta^2}$ and the first-pick prior above, since the negative log-posterior is the Bernoulli log-loss plus the priors' $\sum_t \theta_t^2/(2\sigma_\theta^2)$ and $\gamma_{\mathrm{FP}}^2/(2\cdot 0.5^2)$ terms, which coincide with \eqref{eq:proposed-loss} under the correspondence $\tau \leftrightarrow 1/\sigma_\theta^2$ and $\omega = 1/0.5^2$ (the Bernoulli dispersion being fixed at one, the strength-variance ratio of the Gaussian case reduces to $1/\sigma_\theta^2$ alone). This equivalence is this paper's own construction rather than one carried from Cattelan, Varin, and Firth (\citeyear{cattelan2013dynamic}) or \citet{guan2020analyzing}, and it means $\tau$ plays exactly the role the strength-variance ratio plays in the two-stage model's best linear unbiased prediction (BLUP) shrinkage (\S\ref{sec:twostage}); a team unseen in training receives $\widehat\theta = 0$, the prior mean, under both models. The two tuning constants entering \eqref{eq:onestage-ewma} and \eqref{eq:proposed-loss} are different kinds of knob: $\lambda$ is a covariate-construction parameter that never enters the loss directly, with each grid value rebuilding the columns $E^{\mathrm B}, E^{\mathrm R}$ from scratch, while $\tau$ is the penalty weight inside the loss itself. Both are selected jointly by a grid search on training data alone (\S\ref{sec:workflow}); at fixed $\lambda$ the fit is convex and converges within seconds, so the outer grid is inexpensive to run in full. The criterion surface is shallow in $\lambda$ across the grid and in $\tau$ near the selected value (\S\ref{sec:ablations}).

Two statements place the proposed model within the road map. In mixed-model terms it is the logistic analogue of the two-stage model of \eqref{eq:twostage}, estimated at its maximum a posteriori rather than by REML and BLUP; on the global holdout, where the gap measures, up to the input coding, what a shrunk team strength adds (\S\ref{sec:cattelan}), the completion of Layer~2 scores $0.0121$ $\RPS$ below the purely dynamic benchmark. \textit{In design terms it is deliberately minimal, because the latent update consumes only the binary result: adding richer end-game features produces no statistically clear gain in held-out accuracy (the $d$-ablation of \S\ref{sec:onestage}), a pattern consistent with the strong correlations among the three composite margins (Table~\ref{tab:margincorr}).}

\subsection{The two-stage candidate model} \label{sec:twostage}

The paper's second model is also original to this study. It carries forward the two-stage line described in \S\ref{sec:intro}, in which score-difference models regress a continuous margin on team strengths and the win probability is recovered only afterward \citep{stefani1977football,harville1977use,guan2020analyzing}. It uses the same form design, decay grid, data, and evaluation protocols as the proposed model, so that the parity claim of \S\ref{sec:onestage} is not made against a weak opponent. It is the paper's \emph{second candidate model}, called the two-stage model from here on; ``benchmark'' is reserved throughout for the external references---the classical fits of \S\ref{sec:cattelan} and the market (\S\ref{sec:market}).

The two-stage model's response is a $z$-standardized composite of the game outcome and three end-of-game margins,
\[
Y_i = 0.50\,O_i + 0.25\,\frac{\Delta g_i}{\sigma_g} + 0.15\,\frac{\Delta k_i}{\sigma_k} + 0.10\,\frac{\Delta T_i}{\sigma_T},
\]
built from the blue-minus-red differences in gold, kills, and towers at the end of the map, with weights fixed a priori rather than estimated; \S\ref{sec:ablations} shows similar held-out accuracy across the dense re-weightings tried, a pattern consistent with the margins' strong correlation. Put plainly, the model fit to this composite is a standard sports rating system: it scores winning margins and adjusts each team's rating by how much data that team has accumulated. It is a Gaussian linear mixed model with one stable random strength per team and side-specific form terms of the same design as the proposed model's,
\begin{equation}
\label{eq:twostage}
Y_i = \alpha + \gamma_{\mathrm{FP}}\,(\indic{\text{blue is first-pick}_i}-1)
  + \theta_{b(i)} - \theta_{r(i)}
  + \gamma_1 S^{\mathrm B}_{b(i)}(t_i) - \gamma_2 S^{\mathrm R}_{r(i)}(t_i)
  + \varepsilon_i
\end{equation}
In words, the composite margin of map $i$ is the sum of a global intercept, a first-pick adjustment, the difference between the two teams' stable strengths, and a contribution from each side's recent same-side form, about which the observed margin varies with Gaussian noise. Here $\theta_t \sim \Nrm{0}{\sigma_\theta^2}$ and $\varepsilon_i \sim \Nrm{0}{\sigma^2}$, and $S^{\mathrm B}, S^{\mathrm R}$ follow the same-side win-indicator recursion of \eqref{eq:cattelan-ewma}--\eqref{eq:cattelan-ewma-red}, with per-side decays fixed a priori at $\lambda_1 = \lambda_2 = 0.3$, a value on the same grid the proposed model selects over; it is not tuned (\S\ref{sec:ablations}). Where the first-pick term is estimated, its coefficient carries the prior $\gamma_{\mathrm{FP}} \sim \Nrm{0}{(\sigma/4)^2}$ on the composite scale, the same reference information of sixteen red-first games as in the proposed model; the prior is imposed inside REML by one pseudo-observation, which enters no calibration fit, eligibility rule or reported count. The two-stage model's $S^{\mathrm B}, S^{\mathrm R}$ differ from the Layer-2 restatement of \S\ref{sec:roadmap} in one respect: rather than being initialized at the league-wide prior-season mean, they are zero-initialized at each team's debut, matching the convention adopted for the one-stage family's $E^{\mathrm B}, E^{\mathrm R}$ in \eqref{eq:onestage-ewma} and kept here for comparability between the two architectures. The variance components are estimated by restricted maximum likelihood \citep{patterson1971recovery,harville1977maximum}; the team strengths are predicted by their \textbf{best linear unbiased predictors} \citep{henderson1975best}, shrunk by the reliability weight $\kappa_t = \widehat\sigma_\theta^2 / (\widehat\sigma_\theta^2 + \widehat\sigma^2/n_t)$ toward zero, so that a team unseen in training receives $\widehat\theta = 0$---the same fallback the proposed model's ridge penalty supplies, its correspondence $\tau \leftrightarrow 1/\sigma_\theta^2$ making it this machinery's one-stage analogue. The win probability itself is recovered only in a second step, by a \textbf{Platt map} $\widehat p_i = \operatorname{expit}(c_0 + c_1 \widehat Y_i)$ \citep{platt2000probabilities} fit on the training predictions---a calibration step the proposed model, being fit end-to-end on the win/loss log-loss directly, requires nowhere in its own pipeline (\S\ref{sec:workflow}).

The specification of \eqref{eq:twostage} nests the road map's first two layers as boundary cases rather than as separate competitors. Setting the form coefficients to zero ($\gamma_1 = \gamma_2 = 0$, equivalently the decays $\lambda_1 = \lambda_2 = 0$, which pin the two-stage model's zero-initialized $S^{\mathrm B}, S^{\mathrm R}$ at zero) collapses \eqref{eq:twostage} to a static strength-difference model in the spirit of Layer~1: one random strength per team on the composite response, in the random-effects form of \citet{harville1977use}, with the intercept and first-pick terms retained (the first-pick term enters only where training rows identify it, so it is absent from the holdout fit priced in \S\ref{sec:cattelan}), where \eqref{eq:stefani} as fitted here carries a side-specific pair. Setting the strength variance to zero instead ($\sigma_\theta^2 = 0$, so every BLUP shrinks entirely to the prior mean of zero) yields the Layer-2 dynamic form structure, carrying the linear predictor of \eqref{eq:cattelan-lin} on the margin response rather than the binary outcome, but with the two-stage model's zero-initialization convention for $S^{\mathrm B}, S^{\mathrm R}$ in place of Layer~2's league-mean initialization, so the boundary is exact in structure rather than in every estimated value. Both boundaries are priced on the holdout in \S\ref{sec:cattelan}: the static one exactly, at $0.2286$ (BLUP-shrunk) against $0.2326$ for the least-squares Stefani fit; the dynamic one in its own logistic form, at $0.2382$. The proposed model nests the same two boundaries on the logistic side, with $\tau \to \infty$ forcing $\theta \equiv 0$ and recovering the intercept-augmented form \eqref{eq:cattelan} on the signed input, up to the first-pick term, and $\beta_B = \beta_R = 0$ leaving the static ridge-shrunk strength model alone.

Equation~\eqref{eq:twostage} carries one stable strength per team and fixed form coefficients, and serves as the family's representative throughout. Put simply, the two shrinkage devices place the same Gaussian prior on team strengths (\S\ref{sec:proposed}); what separates them is how the prior's variance is chosen and what else each pipeline carries. Estimating $\sigma_\theta^2$ inside the restricted likelihood requires checking the optimizer's result: one of the three prespecified optimizers reported convergence at points that are not maxima in most of this paper's two-stage fits, so the reported fit is chosen by the qualification rule of \S\ref{sec:workflow}. The selected fits are regular, none lies on the variance boundary, and the qualifying optimizers agree on every single-league fit. Fixing $\tau$ by grid selection instead makes the loss strictly convex in the team strengths, which are therefore identifiable on any slice, with no calibration step; the intercept and slopes still need training rows that identify them, and the first-pick coefficient is kept finite by its prior (\S\ref{sec:proposed}). This point is secondary to the paper's central argument for the design. The two-stage model is compared against the proposed model in \S\ref{sec:onestage} and against the market in \S\ref{sec:market}.

\subsection{Model summary} \label{sec:modelsummary}

Table~\ref{tab:modelsummary} collects, for every model in the road map, the response modeled, the parameters estimated, and the tuning constants set on training data---whether swept over a grid or fixed a priori---rather than estimated by the likelihood, so that shrunk team blocks are never mistaken for free parameters.

\begin{table}[htbp]
\centering
\caption{Model summary across the paired-comparison road map: response actually modeled, parameters actually estimated, and hyperparameters that are ``grid-designed,'' meaning selected \emph{or} fixed on training data and never estimated by the likelihood. In both candidate models the first-pick coefficient, where estimated, carries a prior whose scale is fixed a priori ($\Nrm{0}{0.5^2}$ on the log-odds scale; $\Nrm{0}{(\sigma/4)^2}$ on the composite scale). $T$ is the number of teams ($T = 74$, cross-region pooled). Best linear unbiased predictions and ridge-MAP team blocks are predictions under a prior, not free parameters. The Layer-1 row counts the side-specific adaptation fitted here (\S\ref{sec:roadmap}); Stefani's (\citeyear{stefani1977football}) original assigns one rating per team, $T-1$ of them free with the average fixed. The Layer-2 row counts Cattelan, Varin, and Firth's (\citeyear{cattelan2013dynamic}) own specification, in which the per-side decays are estimated by profile likelihood rather than designed; the re-implementation of \S\ref{sec:cattelan} instead sweeps a single shared $\lambda$ on this paper's grid.}
\label{tab:modelsummary}
\small
\resizebox{\textwidth}{!}{%
\begin{tabular}{llll}
\toprule
Model & Response & Estimated parameters & Grid-designed \\
\midrule
Layer 1 \citep{stefani1977football} & margin $Y_i$ & $2(T{-}1)$ strengths $+$ side level $+\;\sigma^2$ & --- \\
Layer 2 (Cattelan, Varin, and Firth, \citeyear{cattelan2013dynamic}) & binary $O_i$ (logistic) & $\gamma_1,\gamma_2,\lambda_1,\lambda_2$ ($4$; decays by profile ML) & --- \\
Two-stage model \eqref{eq:twostage} & score composite (Gaussian) & $6$ (REML) $+$ Platt $(c_0,c_1)$ & $\lambda_1,\lambda_2$ \\
\midrule
Cattelan re-implementation (\S\ref{sec:cattelan}) & binary $O_i$ (logistic) & $\beta_0,\gamma_1,\gamma_2$ \quad ($3$) & $\lambda$ \\
\textbf{Proposed} \eqref{eq:proposed}, $d{=}1$ & binary $O_i$ (logistic) & $\beta_0,\gamma_{\mathrm{FP}},\beta_B,\beta_R$ ($4$); ridge $\theta$ & $(\lambda,\tau)$ \\
One-stage family \eqref{eq:onestage}, general $d$ & binary $O_i$ (logistic) & $2+2d$ slopes; ridge $\theta$ block & $(\lambda,\tau)$ \\
\bottomrule
\end{tabular}%
}
\end{table}


\section{Data} \label{sec:data}

\subsection{Sources} \label{sec:data:sources}

The modeling corpus is assembled from two publicly accessible sources that share no common join key, and the market benchmark adds a third. \textbf{lolesports} broadcast telemetry supplies the match schedule, per-game metadata including the game-client patch label, and team state recorded at a ten-second cadence---gold, kills, and the map's contested \textbf{objectives}: dragons and barons, neutral monsters whose defeat confers team-wide combat bonuses, and towers and inhibitors, the defensive structures whose destruction opens a path into the opposing base---and supplies the end-of-game margins entering the two-stage model's composite response (\S\ref{sec:twostage}); its endpoints are unofficial, and this paper follows the community API documentation at \texttt{github.com/jpteixeira99/lol-esports-api} (last accessed 24 August 2026). \textbf{Leaguepedia}, a MediaWiki Cargo database, supplies the winning team for $5{,}134$ of the $5{,}135$ modeling games, team names used to check the lolesports side labels, and, from 2026, the first-pick team; its rows are matched to lolesports games on the ten champions picked, with a date check. The remaining winner, of a game whose telemetry ends 46 seconds after its first frame (\S\ref{sec:data:scope}), is recorded from that last frame and confirmed by the series score, which determines it. The corpus reconstruction checks the side labels of every matched game against Leaguepedia's team names, relabeling the games, all of them in the LPL, whose labels contradict those names, and checks every finished series against lolesports' recorded score; one complete series, whose recorded score is contradicted by Leaguepedia and every final frame, keeps Leaguepedia's winners, and series with a game missing are checked for consistency only. Before 2026 the blue side has first pick by rule. \textbf{Polymarket}, a prediction-market API, supplies per-map binary ``Game $N$ winner'' contract prices and is used only as an external forecasting benchmark in \S\ref{sec:market}; it plays no role in fitting any model reported elsewhere in this paper.

\subsection{Volumes and scope} \label{sec:data:scope}

The modeling corpus comprises $5{,}135$ games across six regional leagues---LCK, LPL, LEC, LCS, LCP, and CBLOL---and three international events---First Stand, the Mid-Season Invitational, and Worlds---spanning $74$ teams and $78$ game-client builds over 2024--2026 (counts as of July 2026). Table~\ref{tab:leagues} gives its per-league composition. The historical July collection log recorded $5{,}893$ games and $758$ excluded for missing first-pick assignments; a later reconstruction from the frozen source yields $5{,}457$ collected and $322$ outside the modeling corpus, all of them 2026 games without a first-pick record, mostly from CBLOL, LEC and LCS. They are excluded rather than imputed, which thins those leagues' 2026 test windows and market slices. These collection figures describe different snapshots and are retained as provenance rather than used as analysis denominators. Every game in the $5{,}135$-game modeling corpus has a resolved winner and first-pick assignment.

\begin{table}[htbp]
\centering
\caption{Modeling-corpus game counts by league bucket (first-pick-complete games, $n=5{,}135$; July 2026).}
\label{tab:leagues}
\small
\begin{tabular}{lrlr}
\toprule
League & Games & League & Games \\
\midrule
LPL & 1{,}822 & CBLOL       & 277 \\
LCK & 1{,}276 & LCS         & 216 \\
LEC &   697   & Worlds      & 190 \\
LCP &   432   & MSI         & 157 \\
    &         & First Stand &  68 \\
\bottomrule
\end{tabular}
\end{table}

Two properties of the sources shape the ingestion rules this corpus is built under, and both are stated here because a replicator needs them. The lolesports schedule endpoint is paginated, returning results in pages of roughly eighty events, and must therefore be paged exhaustively: an unpaginated pull silently truncates the corpus rather than failing visibly, and the truncation falls unevenly across leagues, since the affected query windows differ from one league to another, leaving a matchup graph sparser than the true schedule and team-strength estimates correspondingly poorly conditioned---so the defect first presents as an apparently statistical problem rather than as an ingestion error. And the provider's recorded game timestamps are not a reliable start time: they mix two text formats and can differ from the actual start by up to several hours. Every game is therefore timed by its first telemetry frame, which orders games in every protocol, defines which results are completed by a forecast cutoff, and anchors the market price cutoff of \S\ref{sec:market}; anchoring to a later timestamp would sample \emph{settled} prices and produce spuriously perfect market scores. Both rules are load-bearing for every result reported in \S\ref{sec:results}. A third property affects only the telemetry-derived features and completion times. The telemetry endpoint must be walked until each game's final frame, since a fixed horizon silently truncates the longest games, pauses included. For six modeling games the telemetry itself ends before the game does; they keep the values of their last available frame, including the recorded length, and for the four whose telemetry ends before minute 15 the 15-minute features too. Each is counted as completed only at an estimated completion time, Leaguepedia's start plus length plus one minute, or, for the one game Leaguepedia does not list, at the start of its series' next game, which is an upper bound. Leaguepedia's lengths exclude pauses, so a long pause can make such an estimate too early.

\subsection{Evaluation protocol} \label{sec:data:protocol}

Two complementary holdout protocols are used for the main comparisons in this paper, rather than a single design contrasted against an alternative. The \textbf{global time-based split} trains on the oldest $80\%$ of games by calendar time, using only games completed by the split cutoff, and tests on the most recent $20\%$---a single, conservative split whose test window sits entirely after the roster turnover and patch drift the training window predates, so that every static quantity a model estimates is fixed before the test window begins and applied only after it. The \textbf{per-game walk-forward} instead refits at every scored game on games whose results were known by its start time, subject to a minimum training size of $500$ games, and scores $4{,}605$ games in total; this is both the cadence a production deployment of either model would actually run under and the protocol the market comparison of \S\ref{sec:market} inherits, since a market price is itself set before each map is played. Ordinary random cross-validation is invalid for comparing models in this setting and is used nowhere in this paper except in the calibration check of Appendix~\ref{app:calibration}: the exponentially-weighted form covariates central to every dynamic model here (\S\ref{sec:model}) depend on a team's strictly prior games, so a random fold would let future games leak into the very training window used to construct a test game's own features. This subsection specifies \emph{what} is compared under each protocol; the sequence of operations a fit actually executes---feature construction, hyper-parameter selection, refit cadence, and calibration timing---is the workflow specified next.

\subsection{Training and estimation workflow} \label{sec:workflow}

The proposed and two-stage models are specified in \S\ref{sec:proposed} and \S\ref{sec:twostage}; this subsection specifies the sequence of operations a fit of either model runs under either protocol of \S\ref{sec:data:protocol}, kept here as a data-processing and estimation question distinct from the mathematical model definitions themselves. One asymmetry between the two models is a design choice stated here up front rather than left for the reader to discover in the results: the proposed model carries a grid-selection step and no calibration step, while the two-stage model carries the reverse.

\textbf{Feature construction.} Given a training window and a decay ($\lambda$ for the proposed family; per-side $\lambda_1, \lambda_2$ for the two-stage model), the window is walked chronologically and the form states of \S\ref{sec:proposed} and \S\ref{sec:twostage}---the signed-input $E$ states and the win-indicator $S$ states, respectively---are accumulated from each team's debut, zero-initialized as specified in \S\ref{sec:model}; every game's covariate value is its team's form state before the forecast cutoff, updated only by completed games, so no game's own outcome ever enters its own feature. Feature-scaling constants---the two-stage model's margin $z$-scales---are likewise computed from training rows only; the residual and strength variance components are not precomputed here but estimated within the REML fit itself, as described next.

\textbf{Hyper-parameter selection (proposed model only).} The pair $(\lambda, \tau)$ is selected once per protocol, over the grid $\lambda \in \{0.1, 0.2, 0.3, 0.5, 0.7, 0.9\}$ and a $\tau$ grid spanning $0.25$ to $16$, by an inner chronological split of the training data alone: at each grid pair, the model is fit on the first $75\%$ of the training window and scored by log-loss on the remaining $25\%$, and the argmin over the grid is kept. On the global split this inner window is the full training set; under the walk-forward it is the first $500$ games, and the winning pair---together with the feature-scaling constants computed from that same window in the previous step---is then frozen for the remainder of the pass, so that no game the walk-forward ultimately scores ever enters its own hyper-parameter selection. \S\ref{sec:ablations} shows the criterion surface is shallow in $\lambda$ across the grid and in $\tau$ near the selected value. The two-stage model carries no analogous selection step: its decays are fixed a priori at $\lambda_1 = \lambda_2 = 0.3$, a value on the same grid; it is not tuned (\S\ref{sec:ablations}), and its strength-variance shrinkage self-tunes within each fit by restricted maximum likelihood.

\textbf{Fit.} The proposed family is estimated by minimizing the penalized log-loss \eqref{eq:proposed-loss}, a convex problem, on the training window: an L-BFGS-B run followed by Newton steps to a tight gradient tolerance. The two-stage model is estimated by a restricted-maximum-likelihood fit of \eqref{eq:twostage}, with team strengths recovered as best linear unbiased predictors; because optimizers differ in whether they reach a maximum of the restricted likelihood, the reported fit is the highest qualifying restricted-likelihood solution among the prespecified L-BFGS-B, BFGS, and Powell optimizations. A fit qualifies only if the optimizer reports convergence, its likelihood and estimates are finite with a positive residual variance, its random-effect covariance is nonsingular, the Hessian of the restricted log-likelihood is negative definite (a local maximum), and every warning raised about the fit agrees with these checks.

\textbf{Calibration (two-stage model only).} A Platt map is fit on the training window's own predictions---in-sample by design, with no out-of-fold split---and applied to the held-out games that follow. The proposed family requires no analogous step: its probabilities are the direct output of the logistic fit itself, having been fit end-to-end on the win/loss log-loss it is ultimately judged by.

\textbf{Refit cadence.} Under the global split, feature construction, the fit, and calibration each run once. Under the walk-forward, all three rerun at every scored game on games completed by its forecast cutoff (subject to the $500$-game minimum), while hyper-parameter selection does not rerun. The calibration asymmetry between the two models is therefore a matter of in-sample versus out-of-fold estimation, not of once-versus-per-step timing: the two-stage model's Platt map is in fact refit at every walk-forward step, just always on the same rows the mixed model itself was fit on at that step. \S\ref{sec:onestage} quantifies the out-of-fold alternative directly: it moves the two-stage model's score from $0.2274$ to $0.2263$.

\textbf{Computational tools.} The code supporting this study, for data ingestion, corpus correction, estimation, backtesting, verification and figures, was written with the assistance of \textbf{large language models} (Anthropic's Claude), which were also used, under author supervision, to run data audits and draft analyses that the authors then reviewed and re-ran. Claude also drafted manuscript passages and proposed revisions, which the authors reviewed and adopted, revised or rejected; Claude and OpenAI Codex were used to polish manuscript prose under author review. The authors take responsibility for the code, data, analyses and manuscript: every modeling decision and scientific interpretation was made by the authors, and every number reported comes from the authors' pipeline runs.

\textbf{Research ethics and informed consent.} This study analyzes only publicly available professional match records and publicly quoted market prices; it involves no human or animal subjects and no personally identifying data, so no ethical approval was required. Research ethics: Not applicable. Informed consent: Not applicable.


\section{Scoring: the ranked probability score and its reduction to the Brier score} \label{sec:rps}

Forecasts of ordered sporting outcomes are conventionally scored by the \textbf{ranked probability score} (RPS), introduced by \citet{epstein1969scoring} and established as a standard metric for sports-forecast evaluation by \citet{constantinou2012solving}. The RPS is a strictly proper scoring rule for ordered-categorical outcomes \citep{gneiting2007strictly}, meaning a forecaster minimizes its expected value only by reporting its true beliefs, with no incentive to hedge toward or away from any particular category. For a single forecast over $r$ ordered categories with predicted probabilities $p_1, \dots, p_r$ and one-hot outcome indicators $e_1, \dots, e_r$,
\begin{equation}
\label{eq:rps}
\RPS \;=\; \frac{1}{r-1}\sum_{i=1}^{r-1}
          \Bigl(\textstyle\sum_{j=1}^{i} p_j - \sum_{j=1}^{i} e_j\Bigr)^{2}.
\end{equation}
In words, \eqref{eq:rps} accumulates the predicted and realized probability mass up to each ordinal cutoff and penalizes the squared gap between them, so that a forecast placing most of its mass one category away from the true outcome is penalized less than one placing its mass on the opposite end of the ordering---a distinction that matters for genuinely ordered outcomes but collapses, as shown next, for the two-outcome market this paper models throughout.

\subsection{Reduction to the Brier score for the per-map market} The per-map ``Game $N$ winner'' contract has $r=2$ outcomes---blue win or red win---with no ordinal structure to distinguish beyond the two categories themselves. Setting category $1 = $ ``blue win'' with $p_1 = \widehat p$ and $e_1 = O \in \{0,1\}$, \eqref{eq:rps} collapses to a single term,
\begin{equation}
\label{eq:rps2}
\RPS \;=\; \frac{1}{2-1}\,(p_1 - e_1)^2 \;=\; (\widehat p - O)^2 \;=\; \BS,
\end{equation}
the ordinary (single-component) Brier score. Averaged over a holdout set, $\overline{\RPS} = \overline{\BS} = \frac{1}{N}\sum_i (\widehat p_i - O_i)^2$, with $0$ the optimal score and $0.25$ the uninformative baseline achieved by a constant forecast $\widehat p \equiv 0.5$. \textit{Every score reported in \S\ref{sec:results} is therefore simultaneously an RPS and a Brier score, the two being numerically identical for this market;} this paper writes $\RPS$ throughout to match journal convention, reserving the fully general multi-category form \eqref{eq:rps} for the forward-looking remark that closes this section.

\subsection{Standard errors: two conventions, not one} \label{sec:rps:se}

Every score reported below is a mean of per-game losses $s_i = (\widehat p_i - O_i)^2$, so its sampling error is governed by the central limit theorem; two distinct standard errors nonetheless appear throughout \S\ref{sec:results}, and conflating them would understate the precision of every paired comparison this paper reports. The first is the \textbf{marginal plug-in} standard error appropriate to a single forecaster's mean score, estimated by $\mathrm{sd}(s_i)/\sqrt{n}$; conditional on the realized sequence of forecast probabilities $\widehat p_i$ and assuming calibration ($q_i=\widehat p_i$), $\operatorname{Var}(s_i \mid \widehat p_i) = \widehat p_i(1-\widehat p_i)(1-2\widehat p_i)^2 \le 1/16$, so, averaging independent games, the conditional sampling standard error of the mean is at most $1/(4\sqrt n)$. This is only a conditional scale reference rather than either an equality for the realized plug-in estimator or an unconditional bound when forecasts themselves vary across maps. The second, and the one that governs every $\Delta$ reported in this paper, is the \textbf{paired} standard error appropriate to a comparison between two forecasters $A$ and $B$ scored on identical games: writing $d_i = s^A_i - s^B_i$, the estimand is $\Eb[d_i]$ and the correct standard error is the empirical $\mathrm{sd}(d_i)/\sqrt{n}$, with $t = \bar d / \mathrm{SE}$ referred to a standard normal distribution---a Diebold--Mariano test \citep{diebold1995comparing} under squared-error loss and independent games, no forecast horizon here overlapping between games. The direction of each reported $\Delta$ is stated where it appears, since it differs between tables. Because $\operatorname{Var}(d) = \operatorname{Var}(s^A) + \operatorname{Var}(s^B) - 2\rho\,\mathrm{sd}(s^A)\,\mathrm{sd}(s^B)$, the paired standard error scales roughly as $\sqrt{2(1-\rho)}$ times a single marginal standard error, and at the per-game score correlations actually observed in this paper---$\rho = 0.969$ between the proposed and two-stage models in the walk-forward, $\rho = 0.728$ between the proposed model and the market and $\rho = 0.763$ between the two-stage model and the market on the $907$ maps of \S\ref{sec:market}, and $\rho = 0.63$ between the two-stage model and the Cattelan benchmark on the holdout---the paired standard error runs from about the marginal size ($\rho = 0.63$) to roughly four times smaller ($\rho = 0.969$) than the marginal plug-in bound. \textit{The marginal plug-in figure is therefore a conservative scale reference for a single score and never the correct standard error of a difference;} every comparative claim in \S\ref{sec:results}, including the headline parity result between the proposed and two-stage models, uses the paired convention throughout, and per-league rows use the same paired estimator restricted to each league's games, read against a Bonferroni-corrected threshold across the league-level comparisons of the table concerned: six in Table~\ref{tab:perleague}, eight in Table~\ref{tab:market}.

\subsection{Scope for the full ranked probability score} \label{sec:rps:scope}

Because the per-map market reduces exactly to the Brier score, reporting RPS throughout this paper is a matter of nomenclature rather than a distinct analysis. The genuinely multi-category use of \eqref{eq:rps} would arise at the level of the \emph{series} market rather than the map market: a \textbf{best-of-three} (\textbf{Bo3})---the format in which a match ends as soon as one team wins two maps---has an outcome that can be ordered by net game margin, for instance the four-category ordering $\{$team A wins $2$--$0$, A wins $2$--$1$, B wins $1$--$2$, B wins $0$--$2\}$, and \eqref{eq:rps} would then reward forecasts that place their mass ``close'' to the true outcome on that ordinal scale rather than merely on the correct side of it. This paper does not model the series-level market, leaving it as the natural setting in future work where the full ranked probability score does work the Brier score cannot.


\section{Results} \label{sec:results}

All scores reported below are held-out $\RPS$, equal to the Brier score for this two-outcome market by \eqref{eq:rps2}, with lower values indicating better forecasts throughout. The subsections that follow run from league coverage (\S\ref{sec:coverage}) through classical benchmarks (\S\ref{sec:cattelan}) and the architecture comparison (\S\ref{sec:onestage}) to the market (\S\ref{sec:market}), with ablations (\S\ref{sec:ablations}) after the comparison they support. Each forecast uses only results completed by its cutoff: the scored map's start in the global holdout and walk-forward, and the quote time in the market comparison, for both models (\S\ref{sec:market}). The two-stage model is fit throughout with $\lambda_1=\lambda_2=0.3$ and in-sample Platt calibration; the proposed model selects $(\lambda,\tau)$ on training data alone (\S\ref{sec:workflow}).

One protocol detail applies to every holdout table below: a test game is scored only if each team has already appeared on the side it now plays (the seen-on-side rule), so that every model can produce a prediction and the paired tests run on identical games ($115$ of $1{,}027$ cross-region test rows are removed). The filter isolates the unseen-team fallback, $\widehat\theta = 0$ under both architectures.

\subsection{League coverage across the analyses} \label{sec:coverage}

In the primary cross-region analyses of \S\ref{sec:cattelan}--\S\ref{sec:onestage}, every model is fit on the full corpus of \S\ref{sec:data:scope} and evaluated on $912$ held-out games under the global split and on $4{,}605$ games under the per-game walk-forward, in both cases spanning every league in the corpus. The per-league fits in Table~\ref{tab:perleague} instead restrict training and test data to one league at a time, diagnosing performance without cross-region pooling. The matched-market comparison of \S\ref{sec:market} is bounded by data availability: Polymarket began listing per-map LoL markets in October 2025, yielding $912$ matched maps, of which $907$ form the primary market sample (\S\ref{sec:market}). The scores against the market and those on the models' own held-out sets come from different, though overlapping, samples and are not directly comparable.

The proposed model fits and predicts on every reported slice; its fixed ridge penalty makes its team strengths identifiable on single-league designs without a variance-component estimate (\S\ref{sec:twostage}). Table~\ref{tab:perleague} reports six within-league refits, with paired tests read against a Bonferroni cutoff of $0.05/6=0.0083$. LPL favors the two-stage model ($\Delta=+0.0088$, paired SE $0.0025$, $p=0.0005$), whereas LCP favors the proposed model ($\Delta=-0.0462$, paired SE $0.0107$, $p<0.0001$); LCK, LEC, LCS, and CBLOL do not clear the cutoff. The LCP contrast is fragile. All $86$ LCP test games are from 2026, and on them both models score worse than a constant $0.5$ forecast ($0.2725$ and $0.3187$ against $0.25$). In an earlier version of the analysis, which predated the first-pick prior of \S\ref{sec:proposed} and omitted eleven 2025 LCP games whose winners were then unresolved, the comparison had the opposite sign, while the LPL difference was essentially unchanged. Neither single-league result changes the cross-region tie of \S\ref{sec:onestage}. The per-league $(\lambda,\tau)$ selections vary across leagues, consistent with the shallow criterion surfaces documented in \S\ref{sec:ablations}.

\begin{table}[htbp]
\centering
\caption{Per-league comparison of the proposed and two-stage models: within-league fits, with training and test restricted to one league under a $0.2$ time-based split of that league's own games and $(\lambda,\tau)$ selected on its training tail. $\Delta$ is the paired per-game $\RPS$ difference, proposed minus two-stage (negative means the proposed model is ahead, the opposite direction to Table~\ref{tab:onestage}), with its paired standard error in parentheses and a two-sided $p$ against a standard normal (\S\ref{sec:rps:se}). $\Delta$ is computed from unrounded per-game scores, so it need not equal the difference of the rounded $\RPS$ columns. The Bonferroni cutoff for six leagues is $0.0083$. Lower $\RPS$ is better.}
\label{tab:perleague}
\small
\begin{tabular}{lrrrrrr}
\toprule
League & $n_{\text{tr}}$ & $n_{\text{te}}$ & $\RPS_{\text{proposed}}$ & $\RPS_{\text{two-stage}}$ & $\Delta$ (paired SE) & $p$ \\
\midrule
LCK & 1{,}021 & 255 & 0.2157 & 0.2147 & $+0.0011$ ($0.0020$) & $0.60$ \\
LPL & 1{,}458 & 364 & 0.2319 & 0.2230 & $+0.0088$ ($0.0025$) & $0.0005$ \\
LEC & 558   & 139 & 0.2307 & 0.2417 & $-0.0111$ ($0.0090$) & $0.22$ \\
LCS & 173   & 31  & 0.2314 & 0.2350 & $-0.0037$ ($0.0084$) & $0.66$ \\
LCP & 346   & 86  & 0.2725 & 0.3187 & $-0.0462$ ($0.0107$) & $<0.0001$ \\
CBLOL & 222 & 55 & 0.2761 & 0.2525 & $+0.0236$ ($0.0143$) & $0.10$ \\
\bottomrule
\end{tabular}
\end{table}

\subsection{Benchmarks against the classical models} \label{sec:cattelan}

To isolate what a stable team strength buys over a purely dynamic specification, both architectures are benchmarked against the dynamic Bradley--Terry model of Cattelan, Varin, and Firth (\citeyear{cattelan2013dynamic}) fit on the same data---the Layer-2 model of the road map, in exactly the notation of \eqref{eq:cattelan-ewma}--\eqref{eq:cattelan-lin}---and, from the opposite boundary, against the static Stefani line. The re-implementation places a logistic link directly on the binary outcome,
\begin{equation}
\label{eq:cattelan}
P(O_i = 1 \mid \mathcal{F}_{i^-})
   \;=\; \operatorname{expit}\!\bigl(\beta_0 + \gamma_1\, S^{\mathrm B}_{b(i)}(t_i)
         - \gamma_2\, S^{\mathrm R}_{r(i)}(t_i)\bigr),
\end{equation}
conditioning on the history $\mathcal{F}_{i^-}$ of results completed before the forecast cutoff, with $S^{\mathrm B}_{b(i)}, S^{\mathrm R}_{r(i)}$ the same-side win-form EWMAs of \eqref{eq:cattelan-ewma} under a single shared decay $\lambda$. In words, \eqref{eq:cattelan} predicts blue's win probability from the two teams' recent same-side records alone, on the same logistic scale the proposed model uses, with a free intercept carrying the average blue-side advantage and nothing at all carrying either team's persistent quality. The re-implementation departs from Cattelan, Varin, and Firth's (\citeyear{cattelan2013dynamic}) own specification in three deliberate respects: the re-implementation initializes at zero rather than at the league-wide prior-season mean, matching this paper's own models' convention so that it shares the two-stage model's form covariate exactly, an influence that decays geometrically and is in any case covered by the fast-decay end of the $\lambda$ grid; it adds a free intercept $\beta_0$, absorbing the mean blue-side advantage in the role this paper's own global intercepts play elsewhere; and, where the original estimates its per-side smoothing parameters $(\lambda_1, \lambda_2)$ by two-step maximum profile likelihood, the re-implementation sweeps a single shared $\lambda$ across this paper's grid and reports the \emph{best} configuration it attains---an oracle selection made on the evaluation set itself, which favors the benchmark relative to any train-side estimate. The re-implemented abilities remain purely dynamic---no separately estimated stable level, so the form term alone carries the team signal---and the fit is an ordinary logistic regression of $O_i$ on the two side-form EWMAs, the decay profiled on a grid as for this paper's own form terms. At fixed $\lambda$ it estimates three parameters against the proposed model's four plus ridge block and the two-stage model's six (Table~\ref{tab:modelsummary}).

Equation~\eqref{eq:cattelan} is the proposed model \eqref{eq:proposed} without its stable team-strength block and first-pick covariate, on the same link and win/loss response; it codes its form input as the win indicator rather than the signed result, and its $\lambda$ is the evaluation-set best rather than a value selected on training data. On the global holdout the proposed model's first-pick coefficient is fixed at zero (no red-first game in training) and both models sit at $\lambda = 0.1$, so the holdout gap between them measures, up to the input coding, what a shrunk team strength adds to pure dynamics. The two-stage model's gap against \eqref{eq:cattelan} bundles the stable strength with its composite response and Platt map, and the comparisons below do not separate those components.

\begin{table}[htbp]
\centering
\caption{Held-out $\RPS$: the proposed model and the two-stage model against the classical benchmarks---the canonical dynamic Bradley--Terry of Cattelan, Varin, and Firth (\citeyear{cattelan2013dynamic}), swept over $\lambda$, and the two static fits (Stefani's \eqref{eq:stefani} by least squares on the composite response, and the static boundary of \eqref{eq:twostage}, BLUP-shrunk with one strength per team; no form terms, so $\lambda$ is inert)---cross-region, all years, global time-based $0.2$ holdout ($n_{\text{te}}=912$). Lower $\RPS$ is better.}
\label{tab:cattelan}
\small
\begin{tabular}{lrr}
\toprule
Model & $\lambda$ & $\RPS$ \\
\midrule
Cattelan dynamic BT & 0.1 & 0.2382 \\
Cattelan dynamic BT & 0.3 & 0.2410 \\
Cattelan dynamic BT & 0.9 & 0.2477 \\
\midrule
Stefani static \eqref{eq:stefani}, least squares & --- & 0.2326 \\
Static boundary of \eqref{eq:twostage} ($\gamma_1{=}\gamma_2{=}0$, BLUP-shrunk) & --- & 0.2286 \\
\midrule
Two-stage model \eqref{eq:twostage} & 0.3 & 0.2274 \\
\textbf{Proposed} \eqref{eq:proposed} & 0.1 & 0.2261 \\
\bottomrule
\end{tabular}
\end{table}

The proposed model scores $0.2261$ against Cattelan, Varin, and Firth's best configuration ($\lambda=0.1$) at $0.2382$, a gap of $0.0121$, and the two-stage model scores $0.2274$, a gap of $0.0108$ (Table~\ref{tab:cattelan}; both gaps computed from unrounded scores). The re-implemented Cattelan model performs best at the slowest decay on the grid, consistent with its EWMA carrying the team signal without a stable-strength block, whereas both candidate models retain that block. Paired holdout inference gives $\Delta=-0.0108$, two-stage model minus Cattelan, paired SE $0.0049$, $t=-2.21$, and $p=0.03$; the score correlation is $0.63$. On the $4{,}605$-game walk-forward, the corresponding gaps are $0.0107$ for the proposed model and $0.0105$ for the two-stage model. These comparisons include differences beyond stable strength, so they do not isolate one component's causal contribution. On the holdout no first-pick term is active in either candidate model; in the walk-forward that term enters for the games scored from 14 January 2026, and the proposed model runs at $\lambda = 0.9$ against the benchmark's best $\lambda = 0.1$.

\subsubsection{The static classical benchmark} Stefani's static model \eqref{eq:stefani} scores $0.2326$ on the identical holdout; the two-stage model's static boundary, with one REML/BLUP-shrunk strength per team and its intercept retained (its first-pick term, constant in the holdout's training window, drops out), scores $0.2286$. Their $0.0040$ difference combines shrinkage with a change in parametrization. Both score below the purely dynamic Cattelan fit at $0.2382$. The static boundary scores $0.0012$ above the two-stage model ($0.2286$ against $0.2274$), and the dynamic-only fit scores $0.0096$ above the static boundary. The holdout progression is therefore dynamic-only $0.2382$, static-only $0.2286$, both blocks $0.2274$, and the proposed one-stage model $0.2261$. These are descriptive comparisons of fitted specifications, not isolated component effects; of the adjacent steps, only the last is paired-tested in this paper (\S\ref{sec:onestage}), and its difference is not statistically clear.

\subsubsection{Per-game walk-forward corroboration} Under the per-game cadence, each forecast is refit on completed games by its own cutoff. On the common $4{,}605$-game scored set, the proposed model attains $\RPS=0.2208$, the two-stage model $0.2210$, and the best Cattelan configuration $0.2316$ at $\lambda=0.1$. The gaps relative to Cattelan, computed from unrounded scores, are $0.0107$ and $0.0105$, respectively, corroborating the holdout ordering.

\subsection{Architecture head-to-head: the proposed model versus the two-stage model} \label{sec:onestage}

This subsection carries the evidence that selects the proposed model \eqref{eq:proposed} from the general one-stage family \eqref{eq:onestage}: the head-to-head comparison against the two-stage model, and the $d$-ablation that ultimately fixes the latent update's input at the binary result alone. The full feature pool considered, $d=10$, is every directional summary of the previous same-side game available in the telemetry---the win indicator, the three composite margins, inhibitors, barons, dragons, the fifteen-minute gold and kill differentials, and game duration---each an own-minus-opponent quantity within that game and sign-flipped for red, except duration, which has no direction and enters $z$-standardized instead. The family's motivating construction folds these features directly into the latent state: replace the proposed model's binary recursion input with a learned linear combination $w^\top x(t^{-})$ of all such features, and estimate feature weights, side scales, and team strengths jointly against the log-loss of $O_i$. Because the EWMA operator is linear, $\mathrm{EWMA}_\lambda(w^\top x) = w^\top \mathrm{EWMA}_\lambda(x)$, so that model is exactly \eqref{eq:onestage}---a logistic regression on per-feature form EWMAs plus the ridge-penalized signed team block of \eqref{eq:proposed-loss}, an exact reduction rather than an approximation, which is what makes the comparison clean. Three nested variants are fit, with $(\lambda,\tau)$ selected on training data only and the holdout, test filter, and paired protocol identical throughout, on the global $0.2$ time-based holdout of \S\ref{sec:data}: $d=1$ (\texttt{outcome}, exactly the proposed model of \eqref{eq:proposed}), $d=4$ (\texttt{core4}, adding the two-stage model's own three margins, but now learned rather than fixed), and $d=10$ (\texttt{all}, every feature in the pool).

\begin{table}[htbp]
\centering
\caption{One-stage family versus the two-stage model, cross-region all-years holdout ($n_{\text{te}}=912$). $\Delta$ is the paired per-game $\RPS$ difference $\RPS_{\text{two-stage}} - \RPS_{\text{variant}}$ (positive means the variant is nominally ahead), with the Diebold--Mariano standard error of \S\ref{sec:rps:se}. Calibration slopes are held-out (ideal value $1.0$); the one-stage probabilities are native, while the two-stage model carries its in-sample Platt map. The Cattelan row is the external benchmark's best configuration on the same holdout and enters no paired test. Lower $\RPS$ is better.}
\label{tab:onestage}
\small
\begin{tabular}{lrrrrrr}
\toprule
Model & $d$ & $\RPS$ & LogLoss & $\Delta$ vs.\ two-stage & $p$ & Calib.\ slope \\
\midrule
One-stage \texttt{outcome} (proposed) & 1  & 0.2261 & 0.6439 & $+0.0013$ & 0.32 & 0.81 \\
One-stage \texttt{core4}   & 4  & 0.2262 & 0.6450 & $+0.0012$ & 0.44 & 0.79 \\
One-stage \texttt{all}     & 10 & 0.2273 & 0.6475 & $+0.0001$ & 0.95 & 0.76 \\
Two-stage model \eqref{eq:twostage} & --- & 0.2274 & 0.6493 & --- & --- & 0.68 \\
Cattelan dynamic BT ($\lambda{=}0.1$; \S\ref{sec:cattelan}) & --- & 0.2382 & --- & --- & --- & --- \\
\bottomrule
\end{tabular}
\end{table}

\begin{table}[htbp]
\centering
\caption{Association between the binary outcome $O_i$ and the directional (blue$-$red) end-of-game and at-15-minute features, full corpus ($n=5{,}135$): Pearson correlation with $O_i$, single-feature ROC AUC, and the share of blue-win games with a positive value. The three composite margins (gold, kills, towers; marked a) have pairwise correlations of $0.89$--$0.96$; \S\ref{sec:onestage} and \S\ref{sec:ablations} report the related ablations.}
\label{tab:margincorr}
\small
\begin{tabular}{lccc}
\toprule
Feature (blue $-$ red) & Corr.\ with $O_i$ & AUC & \% blue-win, $+$ \\
\midrule
Gold (final)\textsuperscript{a}    & $+0.928$ & $0.999$ & $98.2$ \\
Kills (final)\textsuperscript{a}   & $+0.877$ & $0.993$ & $95.6$ \\
Towers (final)\textsuperscript{a}  & $+0.935$ & $0.999$ & $98.1$ \\
\midrule
Inhibitors (final)     & $+0.854$ & $0.986$ & $96.6$ \\
Barons (final)         & $+0.699$ & $0.900$ & $75.4$ \\
Dragons (final)        & $+0.662$ & $0.885$ & $65.0$ \\
Gold @15               & $+0.478$ & $0.781$ & $72.5$ \\
Kills @15              & $+0.400$ & $0.730$ & $58.9$ \\
\bottomrule
\end{tabular}

\noindent\footnotesize \textsuperscript{a} the three margins entering the two-stage model's composite (\S\ref{sec:twostage}).\normalsize
\end{table}

Five findings organize the evidence in Tables~\ref{tab:onestage} and~\ref{tab:margincorr}. First, the proposed and two-stage models tie on the cross-region global holdout ($0.2261$ versus $0.2274$, paired $p=0.32$) and the $4{,}605$-game walk-forward ($0.2208$ versus $0.2210$, two-stage minus proposed $\Delta=+0.0002$, $p=0.80$). The $d=10$ one-stage variant scores $0.2230$ in the walk-forward, trailing the two-stage model by $0.0019$ (paired SE $0.0010$, $p=0.043$). The paired holdout differences, smaller minus larger dimension, are $-0.0001$ for $d=1$ versus $4$, $-0.0011$ ($p=0.57$) for $d=4$ versus $10$, and $-0.0012$ (SE $0.0016$) end to end. With the margins included, the fitted blue-side win-indicator weight turns negative at $d=10$ ($-0.06$) and is near zero at $d=4$ ($-0.01$; its sign reverses when the six games with incomplete telemetry are excluded); given the margins' correlations we do not interpret individual weights. Second, the holdout feature-dimension ablation is shallow: $d=1,4,10$ score $0.2261$, $0.2262$, and $0.2273$, with no pairwise difference statistically clear ($p=0.45$ to $0.93$), so $d=1$ is chosen on parsimony. The margins' strong correlations in Table~\ref{tab:margincorr} are consistent with limited incremental value, but do not establish its cause. Third, the proposed model produces native probabilities; its holdout calibration slope is $0.81$ against the two-stage model's in-sample-Platt slope of $0.68$, while its full walk-forward intercept is $0.000$ and slope $0.978$. An out-of-fold Platt refit lowers the two-stage holdout score from $0.2274$ to $0.2263$ (Appendix~\ref{app:calibration}). Fourth, predictive rank validity---the Spearman correlation between end-of-training team ranks and test-window win rates over $50$ teams with at least five test games---is $0.448$ for the proposed strengths and $0.458$ for the two-stage BLUPs. Fifth, the cross-region comparison remains close across four equal walk-forward slices and in the market window ($973$ corpus games, $n=963$ scored after ten seen-on-side exclusions), where two-stage minus proposed is $-0.0005$ (paired SE $0.0013$). The proposed model scores $0.2207$ and $0.2208$ at the $\lambda=0.1$ and $0.9$ endpoints, respectively. \textit{We retain the proposed model for its simpler response and lack of a separate calibration step, without claiming a cross-region accuracy advantage over the two-stage model.}

\subsection{Ablations} \label{sec:ablations}

Four design choices are examined by ablation. The proposed model's \textbf{EWMA decay $\lambda$} is relatively insensitive across the tested grid: its walk-forward scores $0.2207$ at $\lambda=0.1$ and $0.2208$ at $\lambda=0.9$ (Appendix~\ref{app:ewma}); the holdout selects $\lambda=0.1$, and the walk-forward selects $\lambda=0.9$. The two-stage model's per-side decays remain fixed a priori at $0.3$ rather than tuned. The \textbf{ridge strength $\tau$} is selected over $0.25$ to $16$; the walk-forward chooses $(\lambda,\tau)=(0.9,4)$ from its first $500$-game training window, and the inner-split log-loss spans $0.0049$ across $\tau\in\{2,4,8\}$ at that decay; the corresponding holdout training-tail spread is under $0.005$. The \textbf{feature dimension $d$} has no statistically clear effect on the global holdout, though $d=10$ trails the two-stage model in the walk-forward (\S\ref{sec:onestage}). For \textbf{response construction}, dense re-weightings of the two-stage composite span $0.0005$ in holdout score; dropping the binary outcome term changes its score from $0.2274$ to $0.2271$, with the largest absolute change among the six within-league refits $0.0028$ in LCP, and replacing the composite by a single margin (gold only or towers only) or by outcome and gold alone changes the holdout score by at most $0.0011$. The two-stage model's held-out accuracy is thus comparatively insensitive to the response constructions tested here.

\subsection{External benchmark: model versus market} \label{sec:market}

\subsubsection{The result} Table~\ref{tab:market} compares both architectures with the market on the primary sample of $907$ of the $912$ matched maps (construction below); both models are scored on the same maps, each forecast made at the map's quote time. The overall comparison includes $134$ tied series-decider maps whose price comes from the series-winner contract; the model prediction remains per-map throughout. Three facts follow. First, the market is ahead even when the analysis is restricted to per-map contracts: model-minus-market $\Delta=+0.011$ ($95\%$ CI $+0.002$ to $+0.019$) for the two-stage model and $+0.011$ ($+0.003$ to $+0.020$) for the proposed model. Second, including tied deciders gives $\Delta=+0.0127$ ($t=+3.14$, $p=0.002$) and $+0.0128$ ($t=+3.2$, $p=0.001$), respectively. Apart from the $14$-map CBLOL slice, the gap is largest at cross-region Worlds, the only league or event slice to clear the Bonferroni $\times8$ cutoff for either model; the LPL and LCK differences are small and not statistically distinguishable from zero. This pattern is consistent with, among other explanations, information available to the market on match day but not to the models, or the difficulty of transferring strength estimates across regions; the comparison does not distinguish them. Third, the architectures still tie on the same $907$ maps (proposed minus two-stage $\Delta=+0.0001$, $p=0.93$); their calibration slopes on this window are $0.77$ for the proposed model and $0.68$ for the two-stage model. The market comparison and the comparison between the two models were substantively unchanged when we retained the map excluded by the preceding-game rule or required a ten-minute buffer after estimated game ends.

\begin{table}[htbp]
\centering
\caption{Market comparison: paired per-map $\RPS$ difference $\Delta=\RPS_{\text{model}}-\RPS_{\text{market}}$ (positive means the market is ahead), with paired $t$, on the primary sample of $907$ of $912$ matched maps, the same maps for both models. The per-map-only row excludes $134$ tied deciders priced from a series-winner contract. League and event rows use a Bonferroni $\times8$ cutoff; CBLOL's $n=14$ is reported for completeness, not interpreted alone.}
\label{tab:market}
\small
\begin{tabular}{lrrrrrr}
\toprule
& \multicolumn{3}{c}{Proposed \eqref{eq:proposed}} & \multicolumn{3}{c}{Two-stage model} \\
League & $n$ & $\Delta$ & $t$ & $n$ & $\Delta$ & $t$ \\
\midrule
LPL         & 316 & $-0.001$ & $-0.08$ & 316 & $-0.004$ & $-0.66$ \\
First Stand & 33  & $+0.008$ & $+0.25$ & 33  & $-0.004$ & $-0.12$ \\
LCK         & 231 & $+0.003$ & $+0.52$ & 231 & $+0.004$ & $+0.75$ \\
LCP         & 149 & $+0.021$ & $+1.95$ & 149 & $+0.030$ & $+2.43$ \\
LEC         & 95  & $+0.029$ & $+2.03$ & 95  & $+0.032$ & $+2.18$ \\
LCS         & 22  & $+0.036$ & $+2.20$ & 22  & $+0.026$ & $+1.57$ \\
Worlds      & 47  & $+0.062$ & $+4.26$ & 47  & $+0.053$ & $+3.49$ \\
CBLOL       & 14  & $+0.100$ & $+1.59$ & 14  & $+0.107$ & $+1.73$ \\
\midrule
Overall     & 907 & $+0.013$ & $+3.24$ & 907 & $+0.013$ & $+3.14$ \\
Per-map only & 773 & $+0.011$ & $+2.63$ & 773 & $+0.011$ & $+2.46$ \\
\bottomrule
\end{tabular}
\end{table}

\begin{figure}[htbp]
\centering
\includegraphics[width=0.95\textwidth]{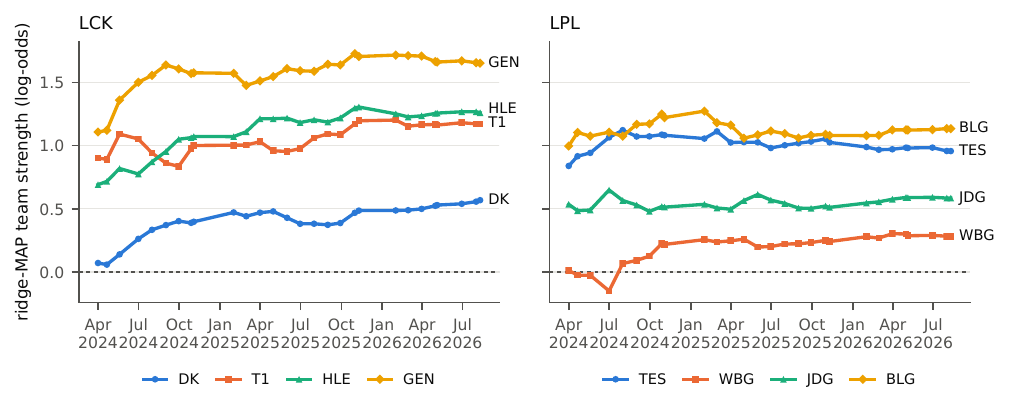}
\caption{Smoothed team-strength trajectories from the proposed model. Each line is the ridge-MAP strength of one team (log-odds scale; zero is the prior mean, dashed) re-estimated on expanding monthly samples by refitting \eqref{eq:proposed}, with each point placed at the last included game's start time, at the walk-forward's frozen pair (\S\ref{sec:ablations}), for the four most-played teams in two leagues: on the left the LCK, with DK (Dplus KIA), T1, HLE (Hanwha Life Esports), and GEN (Gen.G); on the right the LPL, with TES (Top Esports), WBG (Weibo Gaming), JDG (JD Gaming), and BLG (Bilibili Gaming). A line begins at the first monthly checkpoint after the team's debut. Fits use the corpus as of August 2026 ($5{,}233$ games) rather than the frozen corpus behind the tables, so the trajectories are illustrative rather than a reported result; markers distinguish teams in print.}
\label{fig:theta}
\end{figure}

Figure~\ref{fig:theta} traces the proposed model's ridge-MAP team strengths over time, in the spirit of the smoothed-ability plots of Cattelan, Varin, and Firth (\citeyear{cattelan2013dynamic}). The stable block $\theta$ moves slowly and separates teams persistently; the trajectories illustrate the model's fitted strength component rather than explaining the observed market gap.

\begin{figure}[htbp]
\centering
\includegraphics[width=0.95\textwidth]{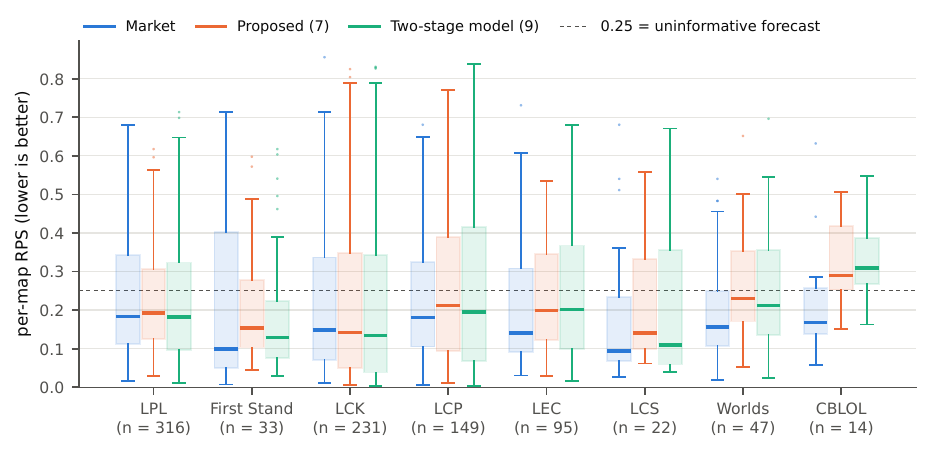}
\caption{Per-map $\RPS$ (the Brier score for this two-outcome contract) by league for the market price, the proposed model \eqref{eq:proposed}, and the two-stage model \eqref{eq:twostage} on the $907$ maps of Table~\ref{tab:market}, the same maps for all three (tied-decider fallback prices included). Boxes span the interquartile range (IQR), with medians marked; whiskers extend to the most extreme observations within $1.5$ IQR of the quartiles, and more extreme points are shown individually. The dashed line at $0.25$ is the score of a constant $0.5$ forecast; lower is better.}
\label{fig:brierleague}
\end{figure}

Figure~\ref{fig:brierleague} shows the per-map score distributions behind Table~\ref{tab:market}'s league and event means. The Worlds panel exhibits the clearest market advantage, whereas the domestic-league distributions overlap substantially; CBLOL's $n=14$ is too sparse to interpret by itself.

\subsubsection{How the market sample is constructed} A map enters the comparison when a Polymarket per-map Game $N$ winner contract matches its teams and game number within the event's dates and has a usable quote before the first telemetry frame, the game's start-time anchor. The contract's two outcomes must name the game's two teams; academy and challengers markets outside the modeling corpus are rejected, and at most one contract is attached to a game, with a same-date slug favored if competing contracts exist. The last pre-start value of the published price series is used without assuming that every value is an order-book midpoint; the quote time is the earlier of the two outcomes' last price points, which always precedes the start. Quotes more than six hours old are rejected ($8$ games), as is an unsettled market; on the accepted quotes, the scored price (the winning outcome's last pre-start value) has a median staleness of $0.5$ minutes and a maximum of $65$ minutes, reached by the one map that the preceding-game rule below removes. A series-winner contract supplies the price only at a tied decider---game 3 of a best-of-three or game 5 of a best-of-five---and only for a game within a day of the fixture date its slug names; $134$ such maps enter the overall sample and are omitted from the per-map-only sensitivity row. For every map both forecasts use the quote-time information set, the results completed by the quote: the two-stage model is refit at the quote, and the proposed model reuses its walk-forward forecast where that information set is the same and is refit at the quote where it differs. A game counts as completed at its start plus its recorded length, an estimate that ignores broadcast lag (for the six games with incomplete telemetry, at the completion time of \S\ref{sec:data:scope}); a preceding game absent from the corpus is timed from Leaguepedia, whose start times are truncated to the minute, with one minute added to its estimated end. Of the $912$ matched maps, the primary sample keeps the $908$ on which both teams have played on their current side by the quote (the seen-on-side rule) and drops one more under the preceding-game rule: for a game after the first in its series, the quote must follow the previous game's estimated end. In the LEC the loser of the previous game holds the right of first selection, and both teams' selections are due within five minutes of that game's end \citep[\S6.8.5]{lec2026rulebook}; the rule applies this premise to every league. One map, quoted before its series had begun, fails the rule, leaving $907$. One author manually traded two of these events (see Competing interests); each recorded fill on them executed against a resting order with that author as taker.

\subsubsection{Scope} The market comparison reported here is a forecasting-quality result only, model $\RPS$ against market-implied $\RPS$: Polymarket functions throughout this paper strictly as an external forecasting benchmark, and nothing built on top of these probabilities is discussed further here.


\section{Conclusion} \label{sec:conclusion}

This paper set out to establish how much modeling machinery a pre-game win-probability forecaster for professional \emph{League of Legends} requires, and how its forecasts compare with a prediction market. The proposed one-stage logistic model combines same-side exponentially weighted form and ridge-shrunk team strength, with a first-pick term only when training data identify it. Across six regional leagues and three international events from 2024--2026, it matches the cross-region accuracy of the independently built two-stage composite mixed model; the latter remains a substantive rival, not a nominal baseline.

A methodological point carries weight beyond this paper's own results. The holdout discipline used throughout is a global time-based split paired with an independent per-game walk-forward, not a contrast between within-season and global protocols, because ordinary random cross-validation is invalid whenever a model's covariates are themselves path-dependent: the exponentially-weighted form terms central to every dynamic model in this paper depend on each team's strictly prior games, so a random fold would let future results leak into the very features used to predict the past. Reporting both protocols together, rather than either alone, is what lets the paper's central parity claim stand on more than a single arbitrary split.

The proposed one-stage model is preferred for parsimony rather than for superior accuracy. It requires neither a composite response nor a separate calibration map, and its calibration slopes are $0.978$ in the walk-forward and $0.81$ on the global holdout (two-stage model: $0.68$ on the holdout); on the global holdout and walk-forward it remains statistically indistinguishable from the two-stage model. Within leagues, LPL favors the two-stage model while LCP favors the proposed model, but the LCP contrast is fragile and both LCP fits score worse than an uninformed constant forecast there (\S\ref{sec:coverage}). Against Polymarket the market is ahead even on per-map contracts, and Worlds is the only league or event slice where the gap survives Bonferroni correction. This pattern is consistent with, among other explanations, information available to the market on match day but not to the models, or the difficulty of transferring strength estimates across regions; the comparison does not distinguish them.

Three limitations qualify these results. In-sample Platt calibration raises the two-stage model's absolute holdout score relative to an out-of-fold refit by $0.0011$ (Appendix~\ref{app:calibration}); this is small in scale but should not be treated as zero when comparing absolute scores. The LCP within-league result rests on $86$ test games and had the opposite sign in an earlier version of the analysis that omitted eleven 2025 games (\S\ref{sec:coverage}), so it cannot establish a general robustness advantage. Finally, the corpus depends on sources with no shared join key and provider conventions that required explicit checks of winners, team sides, first pick, timestamps, and series scores (\S\ref{sec:data:sources}--\S\ref{sec:data:scope}); future extensions must repeat those checks rather than assume the conventions persist.

Three directions are left for future work. A series-level ranked probability score for Bo3 market modeling is the one setting in this paper's own scope where the fully general, multi-category form of \eqref{eq:rps} would do work the Brier score cannot, since a series outcome carries a genuine ordinal structure that a single map's binary result does not. An out-of-fold Platt scheme could be adopted as the two-stage model's default calibration procedure rather than retained only as the robustness check reported in \S\ref{sec:onestage}, closing the one respect in which that model's absolute scores are not directly comparable to the proposed model's own native calibration. And a live, in-game model consuming per-frame telemetry as a match unfolds is a natural downstream consumer of the pre-game forecaster developed here, though building one is explicitly outside this paper's scope, which is confined throughout to information available before a map begins.

\section*{Acknowledgements}

M.-R. G. thanks Teahouse Finance for providing the Claude credits and the
computing resources used for this project.

\section*{Funding}

S.-N. T. is grateful for the financial support from the National Science and
Technology Council of Taiwan under grant 114-2115-M-007-012-MY3,
``Mathematical Foundation of Automated Market Makers.''

\section*{Data and code availability}

The analysis code, the frozen modeling corpus (with the Leaguepedia snapshot used
to build it), the per-map market backtest output including the market
prices used in Section~\ref{sec:market}, and the claim-verification workflow are
held in a private reproducibility repository (release v1.0.11). The analysis
code will be made public under the MIT license no later than publication.
Leaguepedia content is licensed under CC BY-SA 3.0; the Leaguepedia snapshot
will be redistributed under the same license, with attribution. Whether the derived match
data and the Polymarket price data can be redistributed publicly, or provided
to the editor and reviewers under restricted access, is being reconciled with
the providers' terms; the access arrangement will be stated before publication.

\section*{Competing interests}

M.-R. Guan is employed by Teahouse Finance, which provided the Claude credits and
computing resources used for this work. M.-R. Guan has also traded Polymarket's \textit{League of Legends} markets
through a low-balance test account: manually on two events within the period
analyzed in Section~\ref{sec:market} and on five events after it, and, after
that period, through model-driven trading, including use of a version of the
two-stage candidate model of Section~\ref{sec:twostage} and a separate in-game
model. The authors declare no other competing interests.


\appendix
\renewcommand{\thetable}{\thesection.\arabic{table}}
\makeatletter\@addtoreset{table}{section}\makeatother

\section{Calibration: in-sample versus out-of-fold Platt scaling} \label{app:calibration}

The two-stage model's Platt map is fit in-sample throughout \S\ref{sec:results}, on the same training-window predictions as the mixed model of \eqref{eq:twostage} (\S\ref{sec:workflow}); the out-of-fold refit reported in \S\ref{sec:onestage} moves its global-holdout score from $0.2274$ to $0.2263$. The out-of-fold variant assigns training games at random to five folds, refits the mixed model on each fold's complement, predicts each held-out fold, and fits the calibration map once on the pooled predictions before applying it to the outer test set. The folds are drawn with a fixed seed and ignore time order, so a held-out fold's predictions can draw on later training games; this affects only the map's two parameters, every game involved finished before the first test game, and the test-set predictions themselves come from the full-training fit. This reduces the over-separation of in-sample training predictions \citep{niculescumizil2005predicting}. Random folds are used here only to estimate the calibration map, not to compare forecasting models. The $0.0011$ change is small but comparable to the rounded $0.0013$ holdout gap between the two architectures; accordingly, the head-to-head inference in the main text refers to the specified in-sample pipeline, and this diagnostic is reported separately.

\section{EWMA decay grid search} \label{app:ewma}

The proposed model's decay $\lambda$ is selected on training data alone from the six-value grid in \S\ref{sec:workflow}. The frozen-corpus walk-forward scores $0.2207$ and $0.2208$ at the two endpoints, respectively. Table~\ref{tab:lambdasweep} gives a separate six-point diagnostic sweep with $\tau=4$ fixed, run on the August 2026 corpus ($n=4{,}703$ scored games): $\RPS$ spans $0.0004$ and log-loss $0.0010$, computed from unrounded scores. The sweep is a sensitivity diagnostic computed from walk-forward test scores, not a selection step; $\lambda$ is selected on training data alone (\S\ref{sec:workflow}). Its corpus extends the modeling corpus, so only the within-sweep spread is compared with the main results.

\begin{table}[htbp]
\centering
\caption{Diagnostic sweep of EWMA decay $\lambda$ for the proposed model, per-game walk-forward with $\tau=4$ fixed, August 2026 corpus ($n=4{,}703$ scored games); lower is better for both scores. The $\RPS$ spread across the six grid points is $0.0004$. This corpus differs from that behind the main walk-forward results, so its score levels are not used for model selection.}
\label{tab:lambdasweep}
\small
\begin{tabular}{lrrrrrr}
\toprule
$\lambda$ & $0.1$ & $0.2$ & $0.3$ & $0.5$ & $0.7$ & $0.9$ \\
\midrule
$\RPS$  & $0.2211$ & $0.2209$ & $0.2209$ & $0.2210$ & $0.2211$ & $0.2213$ \\
LogLoss & $0.6317$ & $0.6314$ & $0.6314$ & $0.6317$ & $0.6321$ & $0.6324$ \\
\bottomrule
\end{tabular}
\end{table}

No conclusion reported anywhere in this paper depends on the specific decay selected within this range.

\bibliographystyle{plainnat}
\bibliography{references}

\end{document}